\documentclass[twocolumn]{revtex4}
\usepackage{epsfig}
\usepackage{amssymb}
\usepackage{amsmath}
\usepackage{amsfonts}
\usepackage{graphicx}
\usepackage{mathrsfs}
\usepackage[dvips]{color}

\newcommand{\R}{\mathbb{R}}

\newcommand{\fn}{\mathfrak{n}}
\newcommand{\ft}{\mathfrak{t}}
\newcommand{\fr}{\mathfrak{r}}
\newcommand{\fs}{\mathfrak{s}}

\newcommand{\fz}{\mathfrak{z}}

\newcommand{\fK}{\mathfrak{K}}

\newcommand{\bM}{\mathbf{M}}

\newcommand{\be}{\begin{equation}}
\newcommand{\ee}{\end{equation}}
\newcommand{\bea}{\begin{eqnarray}}
\newcommand{\eea}{\end{eqnarray}}
\newcommand{\nn}{\nonumber}

\newcommand{\ed}{\end{document}}

\newcommand{\rg}{{\mbox{\small${\rm G}$}}}
\newcommand{\rk}{{\fK}}

\newcommand{\rz}{{\mbox{\scriptsize${\rm X}$}}}
\newcommand{\rpsi}{{\mbox{\small $\Psi$}}}
\newcommand{\rv}{{\mbox{\small ${\rm V}$}}}

\newcommand{\bi}{\begin{itemize}}
\newcommand{\ei}{\end{itemize}}

\newcommand{\bce}{\begin{center}}
\newcommand{\ece}{\end{center}}

\newcommand{\sE}{\mathscr{E}}

\newcommand{\RE}{{\rm Re}}
\newcommand{\IM}{{\rm Im}}

\begin{document}
\title{Semiclassical Analysis of Spectral Singularities\\ and Their Applications in Optics}
\author{Ali~Mostafazadeh}
\address{Department of Mathematics, Ko\c{c}
University, Sar{\i}yer 34450, Istanbul, Turkey\\
amostafazadeh@ku.edu.tr}

\begin{abstract}

Motivated by possible applications of spectral singularities in
optics, we develop a semiclassical method of computing spectral
singularities. We use this method to examine the spectral
singularities of a planar slab gain medium whose gain coefficient
varies due to the exponential decay of the intensity of pumping beam
inside the medium. For both singly- and doubly-pumped samples, we
obtain universal upper bounds on the decay constant beyond which no
lasing occurs. Furthermore, we show that the dependence of the
wavelength of the spectral singularities on the value of the decay
constant is extremely mild. This is an indication of the stability
of optical spectral singularities.
\medskip

\hspace{6.2cm}{Pacs numbers: 03.65.-w, 03.65.Nk, 42.25.Bs,
24.30.Gd}
\end{abstract}

\maketitle

\section{Introduction}

We say that an $n\times n$ matrix $\mathbf{A}$ has a complete set of
eigenvectors, if we can expand every $n$-dimensional column vector as
a linear combinations of the eigenvectors of $\mathbf{A}$. We use the
same terminology for linear operators acting in infinite-dimensional
Hilbert spaces. In quantum mechanics we make heavy use of the fact that
Hermitian operators have a complete set of eigenvectors. Naturally,
there are non-Hermitian operators that do not have this property. For
non-Hermitian Schr\"odinger operators whose spectrum has a
continuous part a source of the incompleteness of the
eigenfunctions is the presence of what mathematicians
call a spectral singularity. Spectral singularities have been extensively
studied by mathematicians since the 1950's \cite{math}, but remained
essentially unknown to physicists until recently \cite{prl-2009}.

During the past ten years or so it was noticed that some non-Hermitian
operators can be used to serve as the Hamiltonian for a unitary
quantum system provided that one modifies the inner product of the
Hilbert space \cite{review}. Spectral singularities emerge as an
obstruction for the implementation of this Hermitization procedure
for non-Hermitian scattering Hamiltonians. This was initially
noticed in the study of complex point interactions
\cite{jpa-2006b,jpa-2009}. In Ref.~\cite{prl-2009}, we give
the physical meaning of spectral singularities by identifying them
with the energies of certain scattering states that behave exactly
like resonances; they correspond to zero-width resonances
\cite{pra-2009,others}. In \cite{prl-2009,pra-2009} we propose
optical realizations of spectral singularities, and in \cite{p91} we
use a simple toy model to show that the optical spectral singularities
give rise to a lasing effect that takes place exactly at the threshold
gain. The calculation of spectral singularities given in
\cite{prl-2009,pra-2009,p91} rely on the assumption that the gain
coefficient is constant throughout the gain region. This allows for
an essentially exact and analytic treatment of the problem, but it
is practically unattainable. In the present article, we develop a
systematic semiclassical treatment of spectral singularities and
examine its application for more realistic optical systems whose
gain coefficient varies in space.

\begin{figure}
    \begin{center}
    \includegraphics[scale=.55,clip]{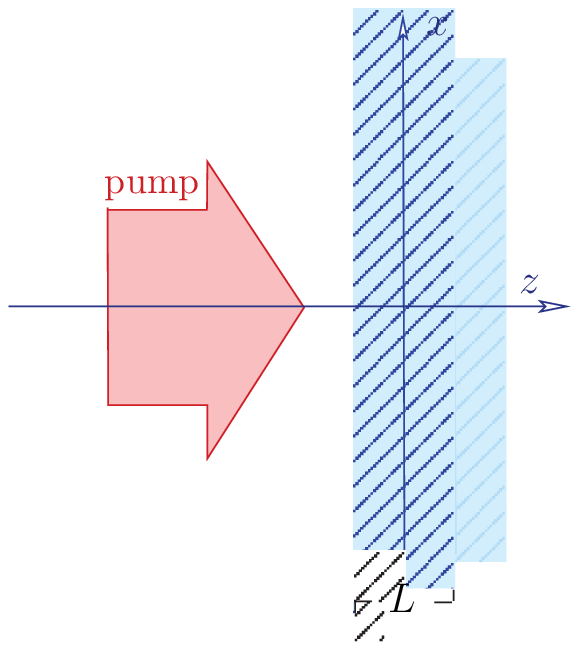}~
    \includegraphics[scale=.55,clip]{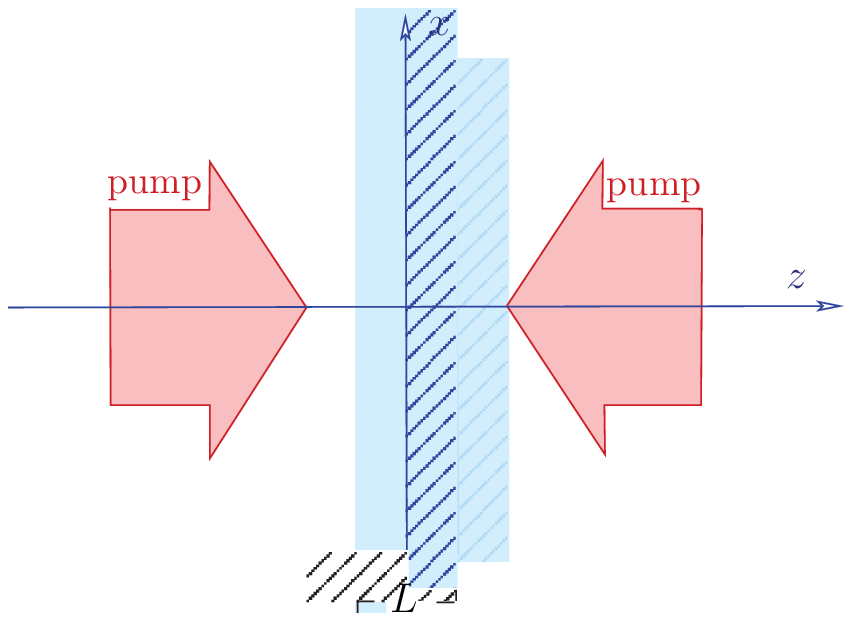}
    \vspace{-.2cm}
    \caption{(Color online) Cross section of an infinite planar slab gain
    medium (dashed blue region) in the $x$-$z$ plane. The arrows represent
    pumping beams. The left- and right-hand figures represent single- and double-pumping, respectively.\label{fig1}}
    \end{center}
    \end{figure}

A simple optical toy model that supports spectral singularities is
an infinite slab gain medium of thickness $L$ that is aligned along
the $x$-$y$ plane \cite{p91}. See figure~\ref{fig1}.
It is easy to check that the following is an exact solution of
Maxwell's equations
    \be
    \vec E(z,t)=E\:e^{-i\omega t}\psi(z)\:\hat e_x,~~~
    \vec B(z,t)=-i \omega^{-1}E\:e^{-i\omega t}\psi'(z)\:\hat e_y,
    \nn
    \ee
where $E$ is a constant, $\hat e_x$ and $\hat e_y$ are the unit
vectors pointing along positive $x$- and $y$-axes, $\psi$ is a
continuously differentiable solution of the Schr\"odinger equation,
     \be
    -\psi''(z)+v(z)\psi(z)=k^2\psi(z),
    \label{sch-eq}
    \ee
$v$ is the potential defined by
    \be
    v(z):=\left\{\begin{array}{ccc}
    k^2[1-\fn(z)^2] &{\rm for}& |z|\leq\mbox{$\frac{L}{2}$},\\ && \\
    0&{\rm for}& |z|>\mbox{$\frac{L}{2}$},\end{array}\right.
    \label{v}
    \ee
$k:=\omega/c$ is the wave number, and $\fn(z)$ is the complex
refractive index of the gain medium.

If the gain medium is obtained by doping a host medium of refraction
index $n_0$ and is modeled as a two-level atomic system with lower and
upper level population densities $N_l$ and $N_u$, resonance frequency
$\omega_0$, and damping coefficient $\gamma$, we have
    \be
    \fn^2= n_0^2-
    \frac{\hat\omega_p^2}{\hat\omega^2-1+i\hat\gamma\,\hat\omega},
    \label{epsilon}
    \ee
where $\hat\omega:=\omega/\omega_0$, $\hat\gamma:=\gamma/\omega_0$,
$\omega_p^2:=(N_l-N_u)e^2/(m_e\varepsilon_0)$, $e$ is electron's
charge, and $m_e$ is its mass. It is not difficult to show that
    \be
    \hat\omega_p^2= 2\hat\gamma\kappa_0\sqrt{n_0^2+\kappa_0^2},~~~~~~~
    \kappa_0:=-\frac{\lambda_0 g_0}{4\pi},
   \label{plasma}
    \ee
where $\lambda_0:=2\pi c/\omega_0$ is the resonance wavelength, and
$g_0$ is the effective gain
coefficient\footnote{By effective gain
coefficient we mean the gain coefficient minus the loss
coefficient.} at the resonance frequency \cite{p91}. For all known
gain media, $\kappa_0\ll n_0$. Therefore,
    \be
    \hat\omega_p^2\approx
    2\hat\gamma n_0\kappa_0=-\frac{\hat\gamma n_0\lambda_0
    g_0}{2\pi}.
     \label{plasma2}
     \ee

In general the gain coefficient $g_0$ is a function of $z$ inside
the gain medium. For example, if we produce the gain by pumping the
medium from the left-hand side, the intensity of the pumping beam
decays exponentially as it penetrates the medium, and we have
    \be
    g_0(z)= (g_\star+\alpha_0)
    \,e^{\mbox{\large $-\frac{L}{2\ell}$}}\:
    e^{\mbox{\large $-\frac{z}{\ell}$}}-\alpha_0
    ~~~~{\rm for}~~~~|z|<\mbox{$\frac{L}{2}$},
    \label{g=1}
    \ee
where $g_\star:=g_0(-L/2)$ is the value of $g_0$ where the pumping
beam enters the gain medium, and $\alpha_0$ is the absorption
(attenuation or loss) coefficient at the resonance frequency. Note
that the maximum gain coefficient $g_\star$ is attained upon
population inversion. Hence $g_\star\leq\alpha_0$, \cite{silfvast}.
If we pump the gain medium from both sides (double-pumping), we find
    \be
     g_0(z)=\left[\frac{g_\star+\alpha_0}{\cosh(\frac{L}{2\ell})}\right]
     \cosh(\mbox{\large$\frac{z}{\ell}$})-\alpha_0
     ~~~~{\rm for}~~~~|z|<\mbox{$\frac{L}{2}$}.
    \label{g=2}
    \ee

It is useful to formulate the problem in terms of the
dimensionless coordinate variable:
    \be
    \rz:=\frac{z}{L}+\frac{1}{2}.
    \label{param0}
    \ee
Then the Schr\"odinger equation~(\ref{sch-eq}) takes the form
    \be
    -\rpsi''(\rz)+\rv(\rz)\rpsi(\rz)=\rk^2\rpsi(\rz),
    \label{sch-eq2}
    \ee
where $\rpsi(\rz):=\psi(L\rz-\mbox{$\frac{L}{2}$})$,
    \bea
    &&
    \rv(\rz):=\left\{\begin{array}{cc}
    \alpha+\beta\, \rg_0(\rz) &{\rm for}~ 0\leq\rz\leq 1,\\
    0&{\rm otherwise} , \end{array}\right.
    \label{param1}\\
    &&\alpha:=\fK^2(1-n_0^2),~~~~~~~~
    \beta:=\frac{n_0\hat\gamma\hat\omega\,\fK}{
    1-\hat\omega^2-i\hat\gamma\hat\omega},
    \label{param2}\\
    &&\rk:=L k=\frac{2\pi L\,\hat\omega}{\lambda_0},
    ~~~~~~
    \rg_0(\rz):=L\,g_0(L\rz-\mbox{$\frac{L}{2}$}).~~~~~~
    \label{param3}
    \eea
We will therefore consider the problem of finding the spectral
singularities of the potentials of the form:
    \be
    \rv(\rz):=\left\{\begin{array}{cc}
    \fz_1+\fz_2\, f(\rz) &{\rm for}~~0\leq\rz\leq 1,\\
    0&{\rm otherwise},\end{array}\right.
    \label{v3}
    \ee
where $f:[0,1]\to\R$ is a piecewise continuous real-valued function,
and $\fz_1$ and $\fz_2$ are complex coupling constants.

\section{Transfer Matrix and Spectral Singularities}

The general solution of (\ref{sch-eq2}) has the form
    \be
    \rpsi(\rz)=\left\{\begin{array}{ccc}
    A_-e^{i\rk\rz}+B_-e^{-i\rk\rz}&{\rm for}&\rz<0,~~~~~\\
    A_0\Phi_1(\rz;\rk)+B_0\Phi_2(\rz;\rk)&{\rm for}&0\leq\rz\leq 1,\\
    A_+e^{i\rk\rz}+B_+e^{-i\rk\rz}&{\rm for}&\rz>1,~~~~~
    \end{array}\right.
    \ee
where $A_0,B_0,A_\pm,B_\pm$ are complex coefficients, and
$\Phi_1(\cdot\,;\,\rk)$ and $\Phi_2(\cdot\,;\,\rk)$ are a pair of
linearly-independent solutions of  (\ref{sch-eq2}) in the interval
$[0,1]$. Imposing the condition that $\rpsi$ and $\rpsi'$ are
continuous at $\rz=0$ and $\rz=1$, we can relate $A_+$ and $B_+$ to
$A_-$ and $B_-$ and determine the transfer matrix $\bM$ of the
system that satisfies: $\vec C_+=\bM\,\vec C_-$ with $\vec
C_\pm:=\left(\begin{array}{c}A_\pm\\B_\pm\end{array}\right)$.

As discussed in \cite{jpa-2009,prl-2009}, the spectral singularities
are the real values of $\rk$ for which the $M_{22}$ entry of the
transfer matrix vanishes. The latter has the form
    \be
    M_{22}=\frac{e^{i\rk}F(\rk)}{2W},
    \label{M22}
    \ee
where
$W:=\Phi_1(\rz;\rk)\Phi'_2(\rz;\rk)-\Phi_1'(\rz;\rk)\Phi_2(\rz;\rk)$
is the Wronskian of $\Phi_1(\cdot\,;\,\rk)$ and
$\Phi_2(\cdot\,;\,\rk)$,
     \be
    F(\rk):=i\rk\Big[\Gamma^-_1(1;\rk)\Gamma^+_2(0;\rk)-
    \Gamma^-_2(1;\rk)\Gamma^+_1(0;\rk)\Big],
    \label{jost}
    \ee
and
    \be
    \Gamma^\pm_j(\rz;\rk):=\Phi_j(\rz;\rk)\pm(i\rk)^{-1}\Phi'_j(\rz;\rk),~~~~
    j=1,2.
    \label{Gammas}
    \ee

Next, we fix the choice of the solutions $\Phi_1(\cdot\,;\,\rk)$ and
$\Phi_2(\cdot\,;\,\rk)$ by demanding that they fulfil the initial
conditions\footnote{This is an acceptable choice, because it gives
$W=i\rk\neq 0$.}:
    \bea
    &&\Phi_1(0,\rk)=1,~~~~~\Phi'_1(0,\rk)=-i\rk,~~~~~\label{special}\\
    &&\Phi_2(0,\rk)=1,~~~~~\Phi'_2(0,\rk)=0.
    \eea
In view of (\ref{Gammas}) and (\ref{special}), we have
$\Gamma_1^+(0;\rk)=0$ and $\Gamma_2^\pm(0;\rk)=1$. These together
with (\ref{M22}) and (\ref{jost}) imply that the spectral singularities of the
potential~(\ref{v3}) are the real zeros of the function
    \be
    F(\rk)=i\rk\,\Phi_1(1;\rk)-\Phi'_1(1;\rk).
    \label{jost2}
    \ee
Equating the right-hand side of this equation to zero and solving
for $\rk$ give
    \be
    \rk=-iG(1,\rk),
    \label{k=}
    \ee
where
    \be
    G(\rz,\rk):=\frac{\Phi'_1(\rz;\rk)}{\Phi_1(\rz;\rk)}=
    \frac{\partial}{\partial\rz} \ln[\Phi_1(\rz;\rk)].
    \label{G=}
    \ee
Because we require $\rk$ to be real, (\ref{k=}) is equivalent to
    \bea
    \RE[G(1,\rk)]=0,
    \label{eq1}\\
    \IM[G(1,\rk)]=\rk.
    \label{eq2}
    \eea
These are the basic real equations that determine the spectral
singularities.

\section{Semiclassical Spectral Singularities}

In order to apply the above procedure of determining spectral singularities of the potentials of  the form (\ref{v3}), we need an explicit expression for the solution $\Phi_1$ of the Schr\"odinger equation~(\ref{sch-eq2}). It is well-known that except for the few exactly solvable special cases, there is no exact and explicit method of constructing such a solution. In this section we employ the method of semiclassical (WKB) approximation to determine $\Phi_1$ and employ the method of Section~2.

The semiclassical solutions of (\ref{sch-eq2}) are given by the
following well-known expression
    \be
    \Psi_\pm(\rz;\rk):=R(\rz;\rk)\,\exp\left[\pm i\int_0^{\rz}
    \sqrt{\rk^2-\rv(\rz)}\,d\rz\right],
    \label{semiclassical}
    \ee
where
    \be
    R(\rz;\rk):=[\rk^2-\rv(\rz)]^{-1/4}.
    \label{R=}
    \ee
These solutions are reliable provided that we can neglect $R''(\rz;\rk)/R(\rz;\rk)$ for all $\rz\in[0,1]$, i.e.,
    \be
    \left|\frac{4[\rk^2-\rv(\rz)]\rv''(\rz)+5\rv'(\rz)^2}{
    16[\rk^2-\rv(\rz)]^3}\right|\ll 1.
    \label{sem-condi}
    \ee

Because $\Phi_1(\,\cdot\,;\rk)$ is a solution of (\ref{sch-eq2}), it must be a linear combination of $\Psi_\pm(\cdot\,;\,\rk)$; there are complex numbers $A$ and $B$ such that
     \be
    \Phi_1(\rz;\rk)=A\,\Psi_+(\rz;\rk)+B\,\Psi_-(\rz;\rk).
    \label{phi1=}
    \ee
We can determine the coefficients $A$ and $B$ by imposing the initial
conditions~(\ref{special}). This gives
    \bea
    A&=&\frac{1}{2R_0}\left[1-\rk R_0^2+
    \frac{i}{4}R_0^6\rv'(0)\right],
    \label{A=}\\
    B&=&\frac{1}{2R_0}\left[1+\,\rk
    R_0^2-\frac{i}{4}R_0^6\rv'(0)\right],~~~~
    \label{B=}
    \eea
where $R_{_\rz}:=R(\rz;\rk)=[\rk^2-\rv(\rz)]^{-1/4}$.

Next, we substitute (\ref{phi1=}) in (\ref{jost2}), equate the
resulting expression to zero, and use (\ref{A=}) and (\ref{B=}) to
derive the following relation for the spectral singularities.
   \bea
  && \exp\left\{2i\!\!\int_0^1\!\!\!\! \sqrt{\rk^2-\rv(\rz)}\,d\rz\right\}=
  \label{ss1}\\
  &&~~~~~\frac{\left[1+\rk R_0^2-\frac{i}{4}\rv'(0)R_0^6\right]
    \left[1+\rk R_1^2+\frac{i}{4}\rv'(1)R_1^6\right]
    }{\left[1-\rk R_0^2+\frac{i}{4}\rv'(0)R_0^6\right]
    \left[1-\rk R_1^2-\frac{i}{4}\rv'(1)R_1^6\right]}.
    \nn
    \eea
For the cases that $\rv(0)=\rv(1)$ and $\rv'(0)=-\rv'(1)$,
(\ref{ss1})  reduces to
    \be
    \exp\left\{2i\!\!\int_0^1\!\!\!\! \sqrt{\rk^2-\rv(\rz)}\,d\rz\right\}=
    \left[\frac{1+\rk R_0^2-\frac{i}{4}\rv'(0)R_0^6
    }{1-\rk R_0^2+\frac{i}{4}\rv'(0)R_0^6}\right]^2.
    \label{ss2}
    \ee
A simple example is the complex barrier potential $V(\rz)$ that has
a constant value $\fz:=\fK^2(1-\fn)$ throughout the interval
$[0,1]$. For this potential, the semiclassical approximation is
exact, and (\ref{ss2}) takes the simple form:
    \be
    e^{2i\rk\, \fn}=\left(\frac{\fn+1}{\fn-1}\right)^2.
    \label{eq10}
    \ee
This coincides with the exact result given in \cite{p91}.

In the remainder of this article we will explore the application of our general results for the potentials of the form~(\ref{v3}). First, we introduce the following pair of  variables:
    \be
    \fr:=\sqrt{1-\frac{\fz_1}{\fK^2}},~~~~\fs:=\frac{\fz_2}{\fK^2-\fz_1}.
    \label{param3n}
    \ee
Using (\ref{v3}) and (\ref{param3n}), we can express (\ref{ss1}) as
    \be
    \exp\left\{2i \fr\fK\int_0^1\sqrt{1-\fs\,f(\rz)}d\rz\right\}=
    \sE(\fK,\fr,\fs),
    \label{ss}
    \ee
where
    \bea
    \sE(\fK,\fr,\fs)&:=& \left[\frac{1+\fr\, p_0(\fs)
    -\fK^{-1}q_0(\fs)}{1-\fr\, p_0(\fs)
    -\fK^{-1}q_0(\fs)}\right]\times\nn\\
    &&~~~~\left[\frac{1+\fr\, p_1(\fs)+\fK^{-1}q_1(\fs)}{
    1-\fr\, p_1(\fs)+\fK^{-1}q_1(\fs)}\right],
    \label{E1}\\
    p_{_\rz}(\fs)&:=&\sqrt{1-\fs f(\rz)},~~~
    q_{_\rz}(\fs):=\frac{i\fs\,f'(\rz)}{4[1-\fs f(\rz)]}.~~~~~~
    \label{param5}
    \eea
Next, we take the logarithm of both sides of (\ref{ss}). In view of
the multi-valuedness of  ``$\ln$'', we can write (\ref{ss}) in the
following equivalent form.
    \be
    \fK=\frac{
    2\pi m+{\rm arg}[\sE(\fK,\fr,\fs)]-i\ln\big|\sE(\fK,\fr,\fs)\big|}{
    2\fr\int_0^1\sqrt{1-\fs\,f(\rz)}d\rz},
    \label{mode}
    \ee
where $m=0,\pm 1,\pm 2,\cdots$ and ``arg[$z$]'' stands for the principal argument of $z$. The integer $m$ that in this way enters into the calculation of spectral
singularities serves as a mode number. This provides a general
explanation for the emergence of a mode number in the study of
spectral singularities.\footnote{In the previous studies of the
subject the emergence of such a mode number could only be linked to
the properties of the specific functions entering the calculations
\cite{prl-2009,pra-2009}.}

Next, we recall that $\fK$ takes real values. Therefore the
right-hand side of (\ref{mode}) must be real. This allows us to
write (\ref{mode}) as the following pair of real equations:
    \bea
    &&\Big\{2\pi m+{\rm arg}[\sE(\fK,\fr,\fs)]\Big\}
    \rho(\fr,\fs)+\nn\\
    &&\hspace{3cm}\ln\big|\sE(\fK,\fr,\fs)\big|\sigma(\fr,\fs)=\fK,
    \label{eq11}\\
    &&\Big\{2\pi m+{\rm arg}[\sE(\fK,\fr,\fs)]\Big\}\sigma(\fr,\fs)
    -\nn\\
    &&\hspace{3cm}\ln\big|\sE(\fK,\fr,\fs)\big|\rho(\fr,\fs)=0,~~~~~~
    \label{eq12}
    \eea
where
    \bea
    \rho(\fr,\fs)&:=&\RE\left[\Big(2\fr\int_0^1
    \sqrt{1-\fs\,f(\rz)}d\rz\Big)^{-1}\right],
    \label{rho-sigma1}\\
    \sigma(\fr,\fs)&:=&\IM\left[\Big(2\fr\int_0^1
    \sqrt{1-\fs\,f(\rz)}d\rz\Big)^{-1}\right].
    \label{rho-sigma2}
    \eea
In view of (\ref{eq12}), we can express (\ref{eq11}) as
    \be
    \fK=\left[\sigma(\fr,\fs)+\frac{\rho(\fr,\fs)^2}{
    \sigma(\fr,\fs)}\right]\ln\big|\sE(\fK,\fr,\fs)\big|.
    \label{K=2}
    \ee

As we will see below, for the typical optical applications,
$|\fs|\ll 1\ll|\fK|$ and $|\fr|\approx n_0$. This observation has
three important consequences. Firstly, it implies that $m$ takes
rather large positive values. Secondly, it confirms the validity of
the semiclassical approximation (\ref{sem-condi}). Thirdly, it
suggests that we can perform a reliable perturbative calculation of
spectral singularities by choosing $\fs$ and $\fK^{-1}$ as
perturbation parameters.\footnote{In a first order perturbative
calculation, in which we ignore the quadratic and higher order terms
in $\fK^{-1}$ and $\fs$, $\fK^{-1}q_0$ and $\fK^{-1}q_1$ drop from
the right-hand side of (\ref{E1}), $\sE$ becomes $\fK$-independent,
and Eqs.~(\ref{eq11}) and (\ref{eq12}) decouple.}

\section{Double-Pumping of an Infinite Slab Gain Medium}

As we pointed out in Section~1, double-pumping of an  infinite slab
gain medium corresponds to a complex potential of the
form~(\ref{v3}) with
    \bea
    &&\fz_1=
    \fK^2\Big[n_0^2(\hat g_\star\ft-1)+1\Big],
    \label{f3b}\\
    &&\fz_2=
    n_0^2\fK^2(\hat g_\star+1)\ft\,,
    \label{f3a}\\
    &&f(\rz)=\frac{\cosh[\nu(\rz-\mbox{$\frac{1}{2}$})]}{
    \cosh(\frac{\nu}{2})}-1,
    \label{f3}
    \eea
where we have used (\ref{g=2}), (\ref{param1}) -- (\ref{param3}), and introduced
    \be
    \hat g_\star:=\frac{g_\star}{\alpha_0}\leq 1,~~~~~
    \ft:=\frac{\hat\gamma\lambda_0\alpha_0}{2\pi n_0
    (1-\omega^2-i\hat\gamma\hat\omega)},~~~~~
    \nu:=\frac{L}{\ell}.
    \ee
Clearly, $\nu=0$ corresponds to the case that the gain coefficient
is uniform throughout the medium. Using (\ref{f3b}) -- (\ref{f3}) in
(\ref{param5}), (\ref{E1}), and (\ref{param3n}), we find
    \bea
    &&\sE(\fK,\fr,\fs)=
    \left[\frac{\fr+1+
    \mbox{\Large$\frac{i\nu\tanh(\mbox{$\frac{\nu}{2}$})\,\fs}{4\fK}
    $}}{\fr-1-\mbox{\Large$\frac{i\nu\tanh(\mbox{$\frac{\nu}{2}$})\,\fs
    }{4\fK}$}}\right]^{2},
    \label{qE=}\\
    &&\fr=n_0\sqrt{1-\hat g_\star\ft},~~~~~~~~
    \fs=\frac{(1+\hat g_\star)\ft}{1-\hat g_\star\ft}.
    \label{r-s=2}
    \eea

Now, consider a typical semi-conductor gain medium \cite{silfvast}
with
    \be
    n_0=3.4,\;\lambda_0=1500\,{\rm
    nm},\;\hat\gamma=0.02,\;\alpha_0=200\,{\rm cm}^{-1},
    \label{values0}
    \ee
    \be
    L\approx 300\,\mu{\rm
    m},~~g_\star\approx 50\,{\rm cm}^{-1},~~~
    \nu\approx 0.1.
    \label{values1}
    \ee
Then we find that at resonance frequency, $\fK=\fK_0:=2\pi
L/\lambda_0>1250$, $|\fs|\approx|\ft|<1.8\times 10^{-3}$, and
$|\fr-n_0|<6.0\times 10^{-4}$. These numerical bounds suggest that
$\fK^{-1}$, $\fs$, and $\ft$ are suitable perturbation parameters
for an accurate perturbative calculation of spectral
singularities.\footnote{Note that $\fK=\hat\omega\fK_0$.} The same
is also true in the vicinity of the resonance frequency,
$\hat\omega\approx 1$, and for other typical gain media where
$\lambda_0/L\ll 1$ and $\lambda_0g_\star\leq\lambda_0\alpha_0\ll 1$.
Furthermore, we can check that for this sample the left-hand side of
(\ref{sem-condi}) is of the order of $10^{-8}$. Therefore,
semiclassical approximation provides an extremely accurate solution
of the problem.

Having related the parameters of the problem with the relevant
physical  data, we wish to investigate the consequences of imposing
(\ref{eq12}) and (\ref{K=2}) that ensure the emergence of a spectral
singularity. Because of the complicated nature of these equations,
this can only be done numerically. It turns out, however, that an
approximate perturbative treatment can play a vital role in
elucidating the physical content of these equations.

\subsection{Perturbation theory and a universal bound on lasing}

In this section we will perform a perturbative treatment of spectral
singularities that involves using $\fK^{-1}$ and $\ft$ as
perturbation parameters. This is a particularly appropriate choice,
because as we stated above $|\fK^{-1}|\approx|\ft|\approx 10^{-3}$.

First we examine the results of the first order perturbation theory
where we ignore quadratic and higher order terms in $\fK^{-1}$and
$\ft$. In particular, because $\fs$ is proportional to $\ft$, we
ignore terms involving $\fs/\fK$ in (\ref{qE=}). This gives
    \bea
    \sE(\fK,\fr,\fs)&\approx&
    \left(\frac{n_0+1}{n_0-1}\right)^{\!\!2}\! \left[1+\frac{2n_0
    \hat g_\star\RE(\ft)}{n_0^2-1}\right]\times\nn\\
    &&~~~~~\exp\left(\frac{2in_0\hat g_\star\IM(\ft)}{n_0^2-1}\right).
    \label{ap-E=1}
    \eea
Note that in this case $\sE(\fK,\fr,\fs)$ does not depend on $\fK$.
As a result, (\ref{eq12}) and (\ref{K=2}) decouple; (\ref{eq12})
determines the location of the spectral singularities, and
(\ref{K=2}) gives the corresponding value of $\fK$.

Next, we use (\ref{rho-sigma1}), (\ref{rho-sigma2}), (\ref{f3}), and (\ref{r-s=2}) to compute $\rho$ and $\sigma$. Again ignoring quadratic and higher
order terms in $\ft$, we find
    \bea
    &&\rho\approx\frac{1+\eta\,\RE(\ft)}{2n_0},~~~~~~~~~~~~~
    \sigma\approx\frac{\eta\,\IM(\ft)}{2n_0},
    \label{ap-rho-sig}
    \eea
where
    \be
    \eta:=\frac{(1+\hat g_\star)\tanh(\frac{\nu}{2})}{\nu}-\frac{1}{2}.
    \label{eta=}
    \ee

It proves useful to examine the case $\hat\omega=1$ separately. In
this case,
    \bea
    &&\ft=\frac{i\lambda_0\alpha_0}{2\pi n_0},~~~~
    \rho\approx\frac{1}{2n_0},~~~~\sigma
    \approx\frac{\lambda_0\alpha_0
    \eta}{4\pi n_0^2},
    \label{ap-rho-sig=}\\
    &&\sE(\fK,\fr,\fs)\approx
    \left(\frac{n_0+1}{n_0-1}\right)^2
     \exp\left[\frac{i\lambda_0g_\star}{\pi(n_0^2-1)}\right].
    \label{ap-E=2}
    \eea
Now, we are in a position to impose (\ref{eq12}) and (\ref{K=2}).
These respectively give
    \bea
    &&\eta\approx\left(\frac{2 n_0}{\lambda_0\alpha_0 m}\right)
    \ln\left(\frac{n_0+1}{n_0-1}\right),
    \label{ap-eta=}\\
    &&\fK_0\approx\frac{\pi m}{n_0}.
    \label{ap-K0=}
    \eea
We can use the latter relation to obtain the mode number for the spectral singularity at resonance wavelength. The result is
    \be
    m\approx \frac{2 n_0 L}{\lambda_0}=1360,
    \label{m=}
    \ee
where we have used $\fK_0=2\pi L/\lambda_0\approx 1256.637$ and the
numerical values (\ref{values0}) and (\ref{values1}).\footnote{In
principal there is no reason for this calculation to yield an
integer value for $m$. Indeed it gives
$m\approx1359.9999999999998$!}

Note that because $\fK_0>0$, (\ref{ap-K0=}) implies  that $\eta>0$.
In light of (\ref{eta=}), this is equivalent to $(\hat
g_\star+1)^{-1}\lessapprox \frac{2}{\nu}\tanh (\frac{\nu}{2})$.
Moreover, we know that $\hat g_\star\leq 1$ and $\frac{2}{\nu}\tanh
(\frac{\nu}{2})\leq 1$. Combining these inequalities, we find
    \be
    \frac{1}{2}\leq \frac{1}{1+\hat g_\star}\lessapprox \frac{2\tanh
     (\frac{\nu}{2})}{\nu}\leq 1.
    \label{ap-condi}
    \ee
In particular, $4\tanh(\frac{\nu}{2})-\nu\geq 0$. This puts
an upper bound on the value of the damping constant $\nu$, namely
$\nu \lessapprox 3.83$. Equivalently, the total damping factor
satisfies
    \be
   1- e^{-\nu}\:\lessapprox~0.978.
    \label{bound}
    \ee
It is remarkable that this bound is independent of other physical
parameters of the system. Note also that because producing a
spectral singularity at the resonance frequency requires smaller
gain than a spectral singularity with a different frequency, this
bound applies more generally for $\hat\omega\approx 1$. More
importantly, in view of the fact that spectral singularities
saturate the laser threshold condition~\cite{p91}, (\ref{bound}) is
actually a universal bound on the possibility of lasing in any
doubly-pumped gain medium.

\begin{figure}
    \begin{center}
    \includegraphics[scale=.40,clip]{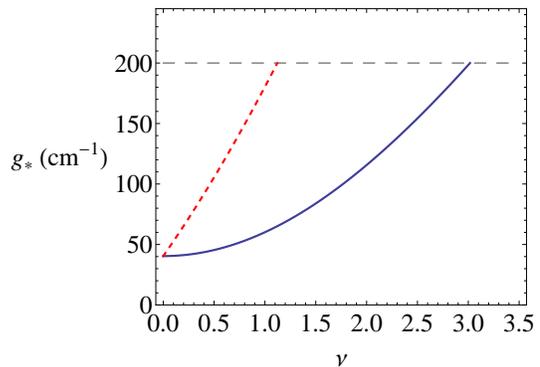}
    \caption{(Color online) A plot of the gain coefficient $g_\star$
    necessary for creating a spectral singularity at the resonance
    wavelength as a function of the damping constant $\nu$ for
    the doubly-pumped sample considered in Section~4 (solid blue curve)
    and singly-pumped sample considered in Section~5 (dashed red curve).
    The dashed grey line marks the upper bound on $g_\star$.\label{fig2}}
    \end{center}
    \end{figure}

Next, we insert (\ref{eta=}) and (\ref{m=}) in (\ref{ap-eta=}) and
solve for $g_\star=\hat g_\star\alpha_0$. This yields
    \be
    g_\star\approx \frac{\nu}{\tanh(\frac{\nu}{2})}\left[\frac{1}{L}
    \ln\left(\frac{n_0+1}{n_0-1}\right)+\frac{\alpha_0}{2}\right]-
    \alpha_0.
    \label{g-star=}
    \ee
Figure~\ref{fig2} shows a plot of the right-hand side of
(\ref{g-star=})  which turns out to be an increasing function of
$\nu$ for $\nu\geq 0$.\footnote{This is actually to be expected,
because it means that for larger values of the decay constant $\nu$
we need larger gain coefficients to maintain the same spectral
singularity.}
In view of the fact that $g_\star$ cannot exceed $\alpha_0$,
(\ref{g-star=}) implies $\nu\lessapprox\nu_{\rm max}$ where
$\nu_{\rm max}$ satisfies
    \be
    2\tanh\left(\frac{\nu_{\rm max}}{2}\right)-\left[\frac{1}{\alpha_0L}
    \ln\left(\frac{n_0+1}{n_0-1}\right)+\frac{1}{2}\right]
    \nu_{\rm max}=0.
    \ee
Solving this equation numerically we find the following improved
bounds.
    \be
    \begin{array}{c}
    \nu\lessapprox\nu_{\rm max}\approx 3.01714112,\\ \\
     1-e^{-\nu}\:\lessapprox 1-e^{-\nu_{\rm max}}\approx 0.9510590651.
     \end{array}
    \label{ap-condi-2}
    \ee
The latter relation means that in order to realize a spectral
singularity  at the resonance frequency, the intensity of each of
the pumping beams should  not drop to less than $e^{-\nu_{\rm
max}}\approx 4.9\%$ of their value in vacuum as they traverse the
gain medium. This corresponds to a $95.1\%$ loss which is much
larger than in typical lasing media.\footnote{Typically the loss is
less than $10\%$.}

Another implication of (\ref{g-star=}) is that in the absence of
damping $(\nu=0)$, the minimum gain coefficient necessary for
creating a spectral singularity (at the resonance frequency) is
given by
    \be
    g_\star=\frac{2}{L}\ln\left(\frac{n_0+1}{n_0-1}\right)\approx
    40.40905~{\rm cm}^{-1}.
    \label{min-g}
    \ee
This is in complete agreement with the results of \cite{p91}.

Next, we recall from \cite{p91} that the presence of a spectral
singularity is extremely sensitive to the values of the parameters
of the system. This suggests that the above first order perturbative
results may not provide a sufficiently accurate description of the
spectral singularities; we need to carry out at least a second order
perturbative calculation in which we also account for the quadratic
terms in $\ft$ and $\fK^{-1}$. Here we summarize the resulting
expressions.
    \bea
    \left|\sE(\fK,\fr,\fs)\right|&\approx&\left(\frac{n_0+1}{n_0-1}\right)^2\Big[
    1+C_1\RE(\ft)+C_2\RE(\ft)^2+\nn\\
    && C_3\IM(\ft)^2+\frac{
    C_4\nu\tanh(\frac{\nu}{2})\IM(\ft)}{\fK}\Big],
    \label{abs-E=}\\
    {\rm arg}[\sE(\fK,\fr,\fs)]&\approx&C_1 \IM(\ft)-2\,C_3\RE(\ft)
    \IM(\ft) \nn\\
    &&-\frac{C_4\nu\tanh(\frac{\nu}{2})\RE(\ft)}{\fK},
    \label{arg-E=}
    \eea
    \bea
    \rho&\approx&\frac{1}{2n_0}\left[1+\eta\RE(\ft)+3\,\xi\,
    \RE(\ft)^2-3\,\xi\,\IM(\ft)^2
    \right],~~~~~\\
    \sigma&\approx&\frac{1}{2n_0}\left[\eta\,\IM(\ft)+6\,\xi\,
    \RE(\ft)\Im(\ft)\right],~~~~
    \label{rho-sigma=20}
    \eea
where
    \bea
    &&C_1:=\frac{2n_o\hat g_\star}{n_0^2-1},~~~~~~
    C_2:=\frac{(3n_0^2+4n_0-1)n_0\hat g_\star^2}{2(n_0^2-1)^2},\nn\\
    &&C_3:=-\frac{n_0(3n_0^2-1)\hat g_\star^2}{2(n_0^2-1)^2},~~~~
    C_4:=-\frac{n_0(1+\hat g_\star)}{n_0^2-1},\nn\\
    &&\xi:=\frac{1}{8}\left\{1+\frac{(1+\hat g_\star)[1+\hat g_\star+
    (\hat g_\star-3)(\frac{\sinh\nu}{\nu})]}{\cosh\nu+1}\right\}.
    \nn
    \eea
Inserting (\ref{abs-E=}) -- (\ref{rho-sigma=20}) in (\ref{eq12}),
noting that $2\pi m$ is of the same order of magnitude as $\fK$, and
keeping the three lowest order terms in the pertubative expansion,
we obtain
    \bea
    &&\pi m\,\eta\,\IM(\ft)-\ln\left(\frac{n_0+1}{n_0-1}\right)+
    6\pi m\,\xi\,\RE(\ft)\IM(\ft)\nn\\
    &&~~~~~-\left[\frac{n_0\hat g_\star}{n_0^2-1}+
    \eta \, \ln\left(\frac{n_0+1}{n_0-1}\right)\right]
    \RE(\ft)\approx 0.
    \label{eq112}
    \eea
 Doing the same for (\ref{eq11}) gives
    \bea
    \fK&\approx&\frac{1}{n_0}\left\{\pi m
    \Big[1+\eta\,\RE(\ft)+3\,\xi\,\RE(\ft)^2-3\,\xi\,\IM(\ft)^2\Big]\right.\nn\\
    &&~~~~~~+\left.
    \left[\frac{n_0\hat g_\star}{n_0^2-1} +
    \eta\ln\left(\frac{n_0+1}{n_0-1}\right)\right]\IM(\ft) \right\}.
    \label{eq111}
    \eea
If we examine the spectral singularity at the  resonance frequency,
we find that (\ref{eq112}) reduces to (\ref{ap-eta=}) while
(\ref{eq111}) yields the following improvement of (\ref{ap-K0=}).
    \be
    \fK_0\approx
    \frac{\pi m}{n_0}+\frac{ \lambda_0\alpha_0}{2\pi n_0^2}\left[
    \left(\eta-\frac{3\xi}{\eta}\right)\ln\left(\frac{n_0+1}{n_0-1}\right)+
    \frac{n_0\hat g_\star}{n_0^2-1}\right].
    \label{ap-K0=2}
    \ee

\subsection{Numerical results}

In this subsection we report the results of a numerical treatment of
Eqs.~(\ref{eq11}) and (\ref{eq12}). This involves fixing the values
of $n_0$, $\lambda_0$, $\hat\gamma$, and $\alpha_0$ as given by
(\ref{values0}), setting $L=300~{\rm nm}$, and determining $\lambda$ and $g_\star$ for various choices of the decay constant $\nu$. Here is a summary of our findings.
    \begin{enumerate}
    \item The numerical results agree with the results of second
    order perturbative calculations at least to 9 significant figures.

    \item It turns out that increasing $\nu$ starting from its minimum
    value $\nu=0$ (unform gain coefficient) has an extremely small effect
    on the wavelength of the spectral singularities. We find spectral
    singularities with almost the same wavelengths but, as expected,
    with larger values of the gain coefficient $g_\star$.
    Table~\ref{table1} lists the values of $m$, $\lambda$ and $g_\star$
    for different $\nu$.
                \begin{table}
                 \begin{center}
                \begin{tabular}{|c||c|}\hline
                ~$m=1335$~ &~$m=1360~$\\
                \begin{tabular}{c|c|c}
                \hline
                $\nu$ & $\lambda~({\rm nm})$ & $g_\star\,({\rm cm}^{-1})$\\ \hline
                0.0 & 1527.6859891 & 175.59110 \\
                0.1 & 1527.6859888 & 175.90413\\
                0.2 & 1527.6859881 & 176.84258\\
                0.3 & 1527.6859868 & 178.40459\\
                0.5 & 1527.6859827 & 183.38565\\
                \end{tabular} &
                \begin{tabular}{c|c|c}
                \hline
                $\nu$ & $\lambda~({\rm nm})$ & $g_\star\,{\rm cm}^{-1}$\\ \hline
                0.0 & 1499.9999833 & 40.40905\\
                0.1 & 1499.9999831 & 40.60936\\
                0.2 & 1499.9999826 & 41.20988\\
                0.3 & 1499.9999819 & 42.20942\\
                0.5 & 1499.9999794 & 45.39683\\
                \end{tabular} \\ \hline
                ~$m=1350$~&~$m=1380~$\\
                 \begin{tabular}{c|c|c}
                 \hline
                $\nu$ & $\lambda~({\rm nm})$ & $g_\star\,({\rm cm}^{-1})$\\
                \hline
                0.0 & 1510.9539613 & 61.80307\\
                0.1 & 1510.9539612 & 62.02123\\
                0.2 & 1510.9539607 & 62.67527\\
                0.3 & 1510.9539598 & 63.76387\\
                0.5 & 1510.9539570& 67.23530\\
                \end{tabular} &
                \begin{tabular}{c|c|c}
                \hline
                $\nu$ & $\lambda~({\rm nm})$ & $g_\star\,({\rm cm}^{-1})$\\
                \hline
                0.0 & 1478.5584532 & 124.17655\\
                0.1 & 1478.5584530 & 124.44660\\
                0.2 & 1478.5584524 & 125.25620\\
                0.3 & 1478.5584514 & 126.60373\\
                0.5 & 1478.5584482 & 130.90084\\
                \end{tabular} \\
                \hline
                \end{tabular}
                \caption{Values of wavelength $\lambda$ and gain
                coefficient $g_\star$  for spectral singularities
                of the doubly-pumped sample considered in Section~4 with
                $m=1335, 1350, 1360, 1380$ for different damping
                coefficients $\nu$.
                \label{table1}}
                \end{center}
                \end{table}

    \item  For $\nu\leq \nu_1\approx 0.22519975$ there is precisely 55
    spectral singularities corresponding to mode numbers
    $m=1333, 1334, \cdots, 1387$. The values of the wavelength $\lambda$
    and gain coefficient $g_\star$ for these spectral singularities are
    depicted in Figure~\ref{fig3}.
    \begin{figure}
    \begin{center}
    \includegraphics[scale=.8,clip]{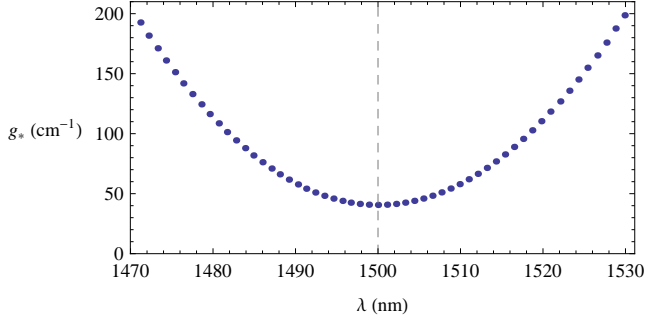}
    \caption{(Color online) Location of the spectral singularities of the
    doubly-pumped sample considered in Section~4 in the
    $g_\star$-$\lambda$ plane for $\nu\leq 0.22519975$. There are
    55 data points corresponds to decreasing values of the mode number
    $m$ from 1387 to 1333. The dashed line marks the resonance
    wavelength $\lambda=\lambda_0=1500~{\rm nm}$. \label{fig3}}
    \end{center}
    \end{figure}
    As expected the spectral singularity with the least amount of gain
    is the one generated at the resonance wavelength.

    \item Figure~\ref{fig3} provides a clear demonstration of how one
    can use spectral singularities to generate a tunable laser
    \cite{p91}. By controlling the intensity of the pumping beam we can
    adjust $g_\star$ to produce lasing at any of the 55 different
    wavelengths shown in Figure~\ref{fig2}. These turns out to be almost
    equally spaced in the range $1471.2$ - $1529.9$~nm with an average
    spacing of $1.07$~nm. 

    \item As one increases $\nu $ beyond $\nu_1\approx 0.22519975$ the
    number of spectral singularities start to decrease. This is because
    in this case in order to create the spectral singularity with mode
    number $m=1335$ or $1387$ the system requires a larger gain
    coefficient $g_\star$ than $\alpha_0=200~{\rm cm}^{-1}$. This is not
    possible, for $\alpha_0$ is the largest value that $g_\star$ can
    take. As one increases $\nu$ further the number of spectral
    singularities keeps dropping. For $\nu> \nu_{\rm max}\approx
    3.01714279$ the last spectral singularity (namely the one at the
    resonance wavelength) cannot be maintained either. This observation
    is in good agreement with the bound: $\nu\lessapprox
    3.01714112$, that we found using the first order perturbative
    calculations.

\end{enumerate}

\section{Single-Pumping of an Infinite Slab Gain Medium}

Consider pumping of the semi-conductor slab gain medium studied in
the preceding section from the left-hand side (See
Figure~\ref{fig1}.) In this case, the gain coefficient $g_0$ and the
parameters $\fz_1$, $\fz_2$, $\fr$ and $\fs$ are given by
(\ref{g=1}), (\ref{f3b}), (\ref{f3a}) and (\ref{r-s=2}),
respectively, while the function $f$ that describes the
space-dependence of the gain coefficient takes the form
    \be
    f(\rz)=e^{-\nu\rz}-1.
    \ee
This in turn implies
    \bea
    \sE(\fK,\fr,\fs)=
    \left(\frac{\fr+1+\mbox{\Large$\frac{i\nu\,\fs}{4\fK}$}}{
                    \fr -1- \mbox{\Large$\frac{i\nu\,\fs}{4\fK}$}}\right) \times
                    ~~~~~~~~~~~~~~~~~~~~~~~~~~\nn\\
    \left(\frac{\fr\sqrt{1+(1-e^{-\nu})\fs}+1-
        \mbox{\Large$\frac{i\nu\,\fs}{4\fK[1+(1-e^{-\nu})\fs]}$}}{
                    \fr\sqrt{1+(1-e^{-\nu})\fs}-1+
        \mbox{\Large$\frac{i\nu\,\fs}{4\fK[1+(1-e^{-\nu})\fs]}$}}\right),
    \label{zE=}
    \eea
where we have used  (\ref{E1}) and (\ref{param5}).

Performing a first order perturbative calculation of $\rho$ and
$\sigma$  yields (\ref{ap-rho-sig}) with $\eta$ given by
    \be
    \eta:=\frac{(1-e^{-\nu})(1+\hat g_\star)}{2\nu}-\frac{1}{2}.
    \label{eta=2}
    \ee
In particular at resonance frequency $\hat\omega=1$, we find that
(\ref{ap-rho-sig=}) still holds but (\ref{ap-E=2}) is slightly
modified:
    \be
    \sE(\fK,\fr,\fs)\approx
    \left(\frac{n_0+1}{n_0-1}\right)^2
     \exp\left[\frac{i\lambda_0g_\star\zeta}{\pi(n_0^2-1)}\right],
    \label{ap-E=22}
    \ee
where $\zeta:=\frac{1}{2}\left[1+e^{-\nu}-{\hat g_\star}^{-1}
(1-e^{-\nu})\right]$. Substituting  (\ref{ap-rho-sig=}) and
(\ref{ap-E=22}) in (\ref{eq12}) and (\ref{K=2}), we recover
(\ref{ap-eta=}) and (\ref{ap-K0=}). Again we can use these equations
and (\ref{eta=2}) to obtain the gain coefficient $g_\star$ as a
function of $\nu$. The result is
    \be
    g_\star\approx\frac{2\nu}{1-e^{-\nu}}\left[\frac{1}{L}
    \ln\left(\frac{n_0+1}{n_0-1}\right)+\frac{\alpha_0}{2}\right]-\alpha_0.
    \label{ap-g=z}
    \ee
The right-hand side of this relation is also an increasing function
of $\nu$. This together with the fact that $g_\star$ cannot exceed
$\alpha_0$ put an upper bound on the allowed values of $\nu$.
Requiring the right-hand side of (\ref{ap-g=z}) not to be larger
than $\alpha_0$ gives
    \be
    \frac{1-e^{-\nu}}{\nu}\gtrapprox \frac{1}{\alpha_0L}
    \ln\left(\frac{n_0+1}{n_0-1}\right)+\frac{1}{2}>\frac{1}{2}.
    \label{g-bound-1}
    \ee
If we enforce the weaker condition, $(1-e^{-\nu})/\nu>1/2$, we find
the following numerical bounds on the decay constant and decay
factor.
    \be
    \nu\lessapprox1.6,~~~~~~1-e^{-\nu}\lessapprox 0.80.
    \ee
Because these are independent of the parameters of the system, they
apply generally for any singly-pumped gain medium. If we enforce the
stronger condition, namely the first inequality in
(\ref{g-bound-1}), we find
    \be
    \begin{array}{c}
    \nu\lessapprox\nu_{\rm max}\approx 1.12208974,\\
    1-e^{-\nu}\lessapprox
    1-e^{-\nu_{\rm max}}\approx 0.67440134.
    \end{array}
    \label{bounds2}
    \ee
These are in extremely good agreement with the (exact) numerical
treatment of spectral singularities that gives
    \be
     \begin{array}{c}
     \nu\lessapprox\nu_{\rm max}\approx 1.12209007,\\
     1-e^{-\nu}\lessapprox
    1-e^{-\nu_{\rm max}}\approx 0.67440144.
    \end{array}
    \label{bounds3}
    \ee
A comparison of these relations with (\ref{ap-condi-2}) shows that
the bounds for the doubly-pumped sample are much weaker than those
on the singly-pumped sample, as it is to be expected.

Another outcome of our numerical investigation is that the
wavelength of the spectral singularities are very close to those
obtained for the doubly-pumped sample of Section~4. However, to
create them one needs higher gain coefficients, particularly as
$\nu$ increases (This is clearly displayed in Figure~\ref{fig2} for
the spectral singularity at the resonance frequency.) Again the
maximum number of spectral singularities that one can create is 55,
and they correspond to mode numbers 1333-1387. All of these can be
created provided that $\nu\leq \nu_1\approx 8.435993\times 10^{-3}$.
This corresponds to a damping of less than $1-e^{-\nu_1}\approx
8.40050985\times 10^{-3}<0.85\%$.

\section{Summary and Conclusion}

In realistic optical models that display spectral singularities the
gain coefficient is a function of space. This motivates the study of
the mathematical problem of finding spectral singularities for
potentials that vanish outside a closed interval. In this article we
identified spectral singularities with real zeros of a particular
complex-valued (so-called Jost) function, derived a semiclassical
expression for this function, and used it to locate the spectral
singularities of a typical semi-conductor gain medium that is
subject to either single- or double-pumping. In both cases, we
performed highly reliable pertutbative calculations and compared
them with the exact numerical results.

The approach pursued here is particularly effective, because it
turns out that for the typical optical realizations of spectral
singularities the semiclassical approximation provides an excellent
description. An important outcome of this approach is the fact that
the inclusion of the effects of the exponential decay of the
intensity of the pumping beams as they pass through the gain medium
does not alter the wavelengths of the spectral singularities
significantly. This seems to be an indication of the stability of
spectral singularities. The mathematical origin of this behavior may
be traced to the fact that spectral singularities are zeros of
complex analytic functions.

An interesting but expected feature of the inclusion of the decay of
the pumping beams is that the gain coefficient $g_\star$ associated
with a given spectral singularity is an increasing function of the
decay parameter $\nu$. This combined with the observation that
$g_\star$ is bounded from above by the absorption coefficient $\alpha_0$
imply the existence of an upper bound $\nu_{\rm max}$ on $\nu$
beyond which no spectral singularity can be produced. This is
actually a bound on any kind of lasing, because spectral
singularities saturate the lasing threshold condition. Using
first-order perturbation theory we obtained a numerical value for
$\nu_{\rm max}$ that is independent of the physical parameters of
the system. Therefore, $\nu_{\rm max}$ is a universal upper bound on
$\nu$ beyond which no lasing occurs.

As one increases $\nu$ starting from $\nu=0$ steadily, one encounter
a number of critical values $\nu_j$ with $j=1,2,\cdots$ at which the
number of allowed spectral singularities drops by one. This
phenomenon may have interesting ramifications for the application of
spectral singularities in producing tunable lasers. For the gain
medium we considered, there are a maximum of 55 spectral
singularities. Double- and single-pumping of this sample
respectively give $0.22519975$ and $0.00843993$ for the first
critical value of $\nu$, namely $\nu_1$. Clearly, $\nu_{\rm
max}=\nu_{55}$.

\vspace{.5cm} \noindent {\em Acknowledgments:} I wish to thank Aref
Mostafazadeh and Ali Serpeng\"{u}zel for illuminating discussions. This work was supported by the Scientific and Technological Research Council of Turkey (T\"UB\.{I}TAK) in the framework of the project no: 110T611 and by Turkish Academy of Sciences (T\"UBA).

\end{document}